\documentclass[onecolumn,12pt]{IEEEtran} 
\usepackage{latexsym,amsmath,amsthm,amssymb,amsxtra,mathrsfs,times,CJK}
\usepackage{color}

\newcommand{\ord}{{\rm ord}}

\newcommand{\tr}{{\rm Tr}}
\newcommand{\gf}{{\rm GF}}

\newcommand{\C}{{\cal C}}

\newcommand{\bc}{{\bf c}}

\newtheorem{theorem}{Theorem}
\newtheorem{lemma}[theorem]{Lemma}

\newtheorem{example}{Example}

\def\BibTeX{{\rm B\kern-.05em{\sc i\kern-.025em b}\kern-.08em
    T\kern-.1667em\lower.7ex\hbox{E}\kern-.125emX}}

\begin{document}

\title{Hamming Weights in Irreducible Cyclic Codes}

\author{
Cunsheng Ding\footnote{C. Ding is with the Department of Computer Science and Engineering,
The Hong Kong University of Science and Technology, Hong Kong, China. Email: cding@ust.hk},
Jing Yang\footnote{J. Yang, the corresponding author, is with the Department of Mathematical Sciences, Tsinghua University,
Beijing, 100084, China. Email: jingyang@math.tsinghua.edu.cn}
}

\date{\today}
\maketitle

\begin{abstract}
Irreducible cyclic codes are an interesting type of codes and have applications
in space communications. They have been studied for decades and a lot of
progress has been made. The objectives of this paper are to survey and extend
earlier results on the weight distributions of irreducible cyclic codes, present
a divisibility theorem and develop bounds on the weights in irreducible cyclic
codes.
\end{abstract}


\section{Introduction}

Throughout this paper, let $p$ be a prime, $q=p^s$ for a positive integer $s$,
and $r=q^m$ for a positive integer $m$. A linear $[n,k,d]$ code over $\gf(q)$
is a $k$-dimensional subspace of $\gf(q)^n$ with minimum (Hamming) distance $d$.
Let $A_i$ denote the number of codewords with Hamming weight $i$ in a code
$\C$ of length $n$. The {\em weight enumerator} of $\C$ is defined by
$$
1+A_1x+A_2x^2+ \cdots + A_nx^n.
$$

A linear $[n,k]$ code $\C$ over the finite field $\gf(q)$ is called {\em cyclic} if
$(c_0,c_1, \cdots, c_{n-1}) \in \C$ implies $(c_{n-1}, c_0, c_1, \cdots, c_{n-2})
\in \C$.
Let $\gcd(n, q)=1$. By identifying any vector $(c_0,c_1, \cdots, c_{n-1}) \in \gf(q)^n$
with
$$
c_0+c_1x+c_2x^2+ \cdots + c_{n-1}x^{n-1} \in \gf(q)[x]/(x^n-1),
$$
any code $\C$ of length $n$ over $\gf(q)$ corresponds a subset of $\gf(q)[x]/(x^n-1)$.
The linear code $\C$ is cyclic if and only if the corresponding subset in $\gf(q)[x]/(x^n-1)$
is an ideal of the ring $\gf(q)[x]/(x^n-1)$.

Note that every idea of $\gf(q)[x]/(x^n-1)$ is principal. Let $\C=(g(x))$ be a
cyclic code. Then $g(x)$ is called the {\em generator polynomial} and
$h(x)=(x^n-1)/g(x)$ is referred to as the {\em parity-check} polynomial of
$\C$.

Let $N>1$ be an integer dividing $r-1$, and put $n=(r-1)/N$.
Let $\alpha$ be a primitive element of $\gf(r)$ and let $\theta=\alpha^N$.
The set
\begin{eqnarray}\label{eqn-irrecode}
\C(r,N) =
 \{(\tr_{r/q}(\beta), \tr_{r/q}(\beta \theta), ...,
              \tr_{r/q}(\beta \theta^{n-1})) :
    \beta \in \gf(r)\}
\end{eqnarray}
is called an {\em irreducible cyclic $[n, m_0]$ code} over $\gf(q)$,
where $\tr_{r/q}$ is the trace function from $\gf(r)$ onto $\gf(q)$,
$m_0$ is the multiplicative order of $q$ modulo $n$ and
$m_0$ divides $m$.

Irreducible
cyclic codes have been an interesting subject of study for many years.
The celebrated Golay code is an irreducible cyclic code and was used
on the Mariner Jupiter-Saturn Mission. They form a special class of codes
and are interesting in theory as they are minimal cyclic codes.
The weight distribution, i.e., the vector $(1, A_1, A_2, \cdots, A_{n-1})$,
of the irreducible cyclic codes has been determined for a small number of
special cases. The objectives of this paper are to survey and extend
earlier results on the weight distributions of irreducible cyclic codes,
present
a divisibility theorem and develop bounds on the weights in irreducible cyclic
codes.

\section{Group characters, cyclotomy, and Gaussian periods}\label{sec-sum}

In this section, we present results on group characters, cyclotomy
and Gaussian sums which will be needed in the sequel.

\subsection{Group characters and Gaussian sums}

Let $\tr_{q/p}$ denote the trace function from $\gf(q)$ to $\gf(p)$.
An {\em additive character} of $\gf(q)$ is a nonzero function $\chi$
from $\gf(q)$ to the set of complex numbers such that
$\chi(x+y)=\chi(x) \chi(y)$ for any pair $(x, y) \in \gf(q)^2$.
For each $b\in \gf(q)$, the function
\begin{eqnarray}\label{dfn-add}
\chi_b(c)=e^{2\pi \sqrt{-1} \tr_{q/p}(bc)/p} \ \ \mbox{ for all }
c\in\gf(q)
\end{eqnarray}
defines an additive character of $\gf(q)$. When $b=0$,
$\chi_0(c)=1 \mbox{ for all } c\in\gf(q),
$
and is called the {\em trivial additive character} of
$\gf(q)$. The character $\chi_1$ in (\ref{dfn-add}) is called the
{\em canonical additive character} of $\gf(q)$.

A {\em multiplicative character} of $\gf(q)$ is a nonzero function
$\psi$ from $\gf(q)^*$ to the set of complex numbers such that
$\psi(xy)=\psi(x)\psi(y)$ for all pairs $(x, y) \in \gf(q)^*
\times \gf(q)^*$.
Let $g$ be a fixed primitive element of $\gf(q)$. For each
$j=0,1,\ldots,q-2$, the function $\psi_j$ with
\begin{eqnarray}\label{dfn-mul}
\psi_j(g^k)=e^{2\pi \sqrt{-1} jk/(q-1)} \ \ \mbox{for } k=0,1,\ldots,q-2
\end{eqnarray}
defines a multiplicative character with order $k$ of $\gf(q)$. When $j=0$,
$ \psi_0(c)=1 \mbox{ for all }
c\in\gf(q)^*,
$
and is called the {\em trivial multiplicative
character} of $\gf(q)$.

Let $q$ be odd and $j=(q-1)/2$ in (\ref{dfn-mul}), we then get a
multiplicative character $\eta$ such that $\eta(c)=1$ if $c$ is
the square of an element and $\eta(c)=-1$ otherwise. This $\eta$
is called the {\em quadratic character} of $\gf(q)$.

Let $\psi$ be a multiplicative character with order $k$ where $k|(q-1)$ and $\chi$ an additive character of
$\gf(q)$. Then the {\em Gaussian sum} $G(\psi,\chi)$ of order $k$ is defined by
\begin{eqnarray}
G(\psi,\chi)=\sum_{c\in\gf(q)^*} \psi(c)\chi(c). \nonumber
\end{eqnarray}

Since $G(\psi,\chi_b)=\bar{\psi}(b)G(\psi,\chi_1)$, we just consider $G(\psi,\chi_1)$, briefly denoted as $G(\psi)$, in the sequel. If $\psi\neq\psi_0$
, then
\begin{equation}\label{equ-Gvalue}
|G(\psi)|=q^{1/2}.
\end{equation}

Generally, to explicitly determine the value of Gaussian sums is a challenging task.
At present, they can be determined in a few cases. Among them is the following case of $k=2$.

If $q=p^s$, where $p$ is an odd prime and $s$ is a positive integer, then
\begin{eqnarray}\label{eqn-etachi1}
G(\eta)=
\left\{
\begin{array}{ll}
(-1)^{s-1}q^{1/2}&\mbox{if $p\equiv 1 \pmod{4}$}, \\
(-1)^{s-1}(\sqrt{-1})^s q^{1/2}&\mbox{if $p\equiv 3 \pmod{4}$}.
\end{array}
\right.
\end{eqnarray}

The following result (\cite{HaNi83}) is useful in the sequel.

\begin{lemma}\label{lem-quadratic}
Let $\chi$ be a nontrivial additive character of $\gf(q)$ with $q$
odd, and let $f(x)=a_2x^2+a_1x+a_0\in\gf(q)[x]$ with $a_2\neq 0$.
Then
\begin{eqnarray}\label{eqn-weil1}
\sum_{c\in\gf(q)}\chi(f(c))=\chi(a_0-a_1^2(4a_2)^{-1})\eta(a_2)G(\eta).
\end{eqnarray}
\end{lemma}

The Gaussian sums of small order, such as $k=3$, 4, 5, 6, and 12, can be also determined, see \cite{BEW}. In another special case, called ``semi-primitive'' case, the  Gaussian sums are known and given in
the following two lemmas \cite{BEW}.

\begin{lemma}\label{lemma1-semi}
Assume that $N\neq2$ and there exists a positive integer $j$ such that $p^j \equiv -1 \pmod{N}$, and
the $j$ is the least such. Let $q=p^{2j\gamma}$ for some integer $\gamma$. Then the Gaussian sums of order $N$ over $\gf(q)$ are given by
\[G(\psi)=\left\{\begin{array}{ll}
(-1)^{\gamma-1}\,\sqrt{q},&\mbox{if}\,~p=2,\\
(-1)^{\gamma-1+\frac{\gamma(p^j+1)}{N}}\,\sqrt{q},&\mbox{if}\,~p\geqslant 3.\end{array}\right.\]
\end{lemma}

\begin{lemma}\label{lemma2-semi}
Let notations be defined as in Lemma \ref{lemma1-semi}. For $1\leqslant i\leqslant N-1$, the Gaussian sums $G(\psi^i)$ are given by
\[G(\psi^i)=\left\{\begin{array}{ll}
(-1)^{i}\,\sqrt{q},&\mbox{if $N$ is even, $p,\gamma$ and $\frac{p^j+1}{N}$ are odd};\\
(-1)^{\gamma-1}\,\sqrt{q},&\mbox{otherwise.}\end{array}\right.\]
\end{lemma}

If $p$ generates a subgroup of group $(\mathbb{Z}/N\mathbb{Z})^*$ with index $[(\mathbb{Z}/N\mathbb{Z})^*:\langle p\rangle]=2$ and $-1\not\in\langle p\rangle \subset(\mathbb{Z}/N\mathbb{Z})^*$, which is the so-called ``{\em quadratic residues}" or ``{\em index 2}" case, Gaussian sums are also explicitly determined. See \cite{Y-X10} and its references for details. We list one of the results \cite{Y-X10} in the index 2 case below, which is useful in the sequel.

\begin{lemma}\label{lemma-index2}
Let $N_1=l^\lambda$ where $3\neq l\equiv 3\pmod{4}$ is a prime and $\lambda$ is a positive integer. Let $f=\ord_{N_1}(p),\ r=p^{fs}$ for some positive integer $s$, and \(\psi\) be a primitive multiplicative character of order $N_1$ over ${\gf}(r)^*$. Assume that $f=\frac{\varphi(N_1)}{2}$, which means that $p$ generates the quadratic residues modulo $N_1$, then, for $1\leqslant t\leqslant \lambda$, we have that
$$\begin{array}{rl}
G(\psi^{\lambda-t})&=(-1)^{s-1}\cdot p^{\frac{s(f-hl^{\lambda-t})}{2}}\cdot \left({\frac{a+b\sqrt{-l}}{2}} \right)^{sl^{\lambda-t}}\\
    &:=P_t^{(s,\lambda)}\left(A_t^{(s,\lambda)}+B_t^{(s,\lambda)}\sqrt{-l}\right),
\end{array}$$
where $h$ is the ideal class number of $\mathbb{Q}(\sqrt{-l})$, the
integers $a, b$ are given by
$$\left\{ {\begin{array}{l}
 a^2+lb^2=4p^h\\
 a\equiv -2p^{\frac{l-1+2h}{4}}\pmod{l},
\end{array}} \right.$$
and $P_t^{(s,\lambda)},~A_t^{(s,\lambda)},~B_t^{(s,\lambda)}\in\mathbb{Z}$ are defined as
\begin{equation}\label{equ1-index2}
    \begin{array}{l}P_t^{(s,\lambda)}=(-1)^{s-1}\cdot p^{\frac{s(f-hl^{\lambda-t})}{2}};\\
    A_t^{(s,\lambda)}={\rm Re}\left({\frac{a+b\sqrt{-l}}{2}} \right)^{sl^{\lambda-t}};\ B_t^{(s,\lambda)}={\rm Im}\left({\frac{a+b\sqrt{-l}}{2}} \right)^{sl^{\lambda-t}}\Big/\sqrt{l}.\end{array}
\end{equation}
\end{lemma}

\subsection{Cyclotomy}

Let $r-1=nN$ for two positive integers $n>1$ and $N>1$, and let
$\alpha$ be a fixed primitive element of $\gf(r)$.
Define $C_{i}^{(N,r)}=\alpha^i \langle \alpha^{N} \rangle$ for $i=0,1,...,N-1$, where
$\langle \alpha^{N} \rangle$ denotes the
subgroup of $\gf(r)^*$ generated by $\alpha^{N}$. The cosets $C_{i}^{(N,r)}$ are
called the {\em cyclotomic classes} of order $N$ in $\gf(r)$.
The {\em cyclotomic numbers} of order $N$ are
defined by
\begin{eqnarray*}
(i, j)^{(N,r)}=\left|(C_{i}^{(N,r)}+1) \cap C_{j}^{(N,r)}\right|
\end{eqnarray*}
for all $0 \leq i \leq  N-1$ and $0 \leq j \leq  N-1$.

We will need the following lemma (\cite{DY}) in the sequel.

\begin{lemma}\label{lem-CycloSum}
Let $r-1=nN$ and let $q$ be a prime power. Then
\[
     \sum_{u=0}^{N-1} (u,u+k)^{(N,r)} =\left\{\begin{array}{ll}
                          n-1, & \mbox{ if $k=0$}, \\
                           n,  & \mbox{ if $k \neq 0$}.\\
			\end{array}
			\right.
\]
\end{lemma}

\vspace{.2cm}
To determine the weight distribution of some classes of linear codes in the sequel, we need the
following lemma.

\begin{lemma}\label{lem-haw}
Let $e_1$ be a positive divisor of $r-1$
and let $i$ be any integer with $0 \le i <e_1$.
We have the following multiset equality:
\begin{eqnarray}\label{eqn-fgfg0}
\left\{xy: y \in \gf(q)^*, \ x \in C_i^{(e_1,r)}\right\} = \frac{(q-1)\gcd((r-1)/(q-1), e_1)}{e_1}*C_i^{(\gcd((r-1)/(q-1),e_1),r)},
\end{eqnarray}
where $\frac{(q-1)\gcd((r-1)/(q-1), e_1)}{e_1}*C_i^{(\gcd((r-1)/(q-1),e_1),r)}$ denotes the multiset
in which each element in
the set $C_i^{(\gcd((r-1)/(q-1),e_1),r)}$ appears in the multiset with multiplicity $\frac{(q-1)\gcd((r-1)/(q-1), e_1)}{e_1}$.
\end{lemma}

\begin{proof}
We need to prove the conclusion for $i=0$ only because
$$C_i^{(\gcd((r-1)/(q-1),e_1),r)}=\alpha^i C_0^{(\gcd((r-1)/(q-1),e_1),r)}.$$
Note that every $y \in \gf(q)^*$ can be expressed as $y=\alpha^{\frac{r-1}{q-1}\ell}$
for an unique $\ell$ with $0 \le \ell < q-1$ and every $x \in C_0^{(e_1,r)}$ can be expressed as
$x=\alpha^{e_1j}$ for an unique $j$ with $0 \le j < (r-1)/e_1$. Then
we have
$$
xy=\alpha^{\frac{r-1}{q-1}\ell + e_1j}.
$$
It follows that
$$
xy=\alpha^{\frac{r-1}{q-1}\ell + e_1j}=
(\alpha^{\gcd((r-1)/(q-1),e_1)})^{ \frac{r-1}{(q-1)\gcd((r-1)/(q-1),e_1)}\ell + \frac{e_1}{\gcd((r-1)/(q-1),e_1)}j}.
$$
Note that
$$
\gcd\left( \frac{r-1}{(q-1)\gcd((r-1)/(q-1),e_1)}, \frac{e_1}{\gcd((r-1)/(q-1),e_1)}\right)=1.
$$
When $\ell$ ranges over $0 \le \ell < q-1$ and $j$ ranges over $0 \le j < (r-1)/e_1$,
$xy$ takes on the value $1$ exactly $\frac{q-1}{e_1}\gcd((r-1)/(q-1), e_1)$ times.

Let $x_{i_1} \in C_0^{(e_1,r)}$ for $i_1=1$ and $i_1=2$, and let $y_{i_2} \in \gf(q)^*$ for $i_2=1$ and $i_2=2$.
Then $\frac{x_1}{x_2} \in C_0^{(e_1,r)}$ and $\frac{y_1}{y_2} \in \gf(q)^*$. Note that
$x_1y_1=x_2y_2$ if and only if $\frac{x_1}{x_2} \frac{y_1}{y_2} =1$. Then the
conclusion of the lemma for the case $i=0$ follows from the discussions above.
\end{proof}

\subsection{Gaussian periods}

\vspace{.2cm}
The {\em Gaussian periods} are defined by
$$
\eta_i^{(N,r)} =\sum_{x \in C_i^{(N,r)}} \chi(x), \quad i=0,1,..., N-1,
$$
where $\chi$ is the canonical additive character of $\gf(r)$.

The following lemma presents some basic properties of Gaussian periods, and will be
employed later.

\begin{lemma}\label{lem-periodpro}  \cite{Stor}
Let symbols be the same as before. Then we have
\begin{enumerate}
\item $\sum_{i=0}^{N-1} \eta_i = -1.$
\item $\sum_{i=0}^{N-1} \eta_i \eta_{i+k} = r\theta_k-n$ for all $k \in \{0,1, \cdots, N-1\}$, where
$$
\theta_k=\left\{ \begin{array}{ll}
                          1 & \mbox{ if $n$ is even and $k=0$} \\
                          1 & \mbox{ if $n$ is odd and $k=N/2$} \\
                          0 & \mbox{ otherwise,}
                          \end{array}
                          \right.
$$
\end{enumerate}
and equivalently $\theta_k=1$ if and only if $-1 \in C_{k}^{(N,r)}$.
\end{lemma}

Gaussian periods are closely related to Gaussian sums. By the discrete Fourier transform, it is known that
\begin{equation}\label{equ-G sum& G period}
    \eta_i^{(N,r)}=\frac{1}{N}\sum\limits_{j=0}^{N-1}\zeta_N^{-ij}G(\psi^j)=\frac{1}{N}\left[-1+\sum\limits_{j=1}^{N-1}\zeta_N^{-ij}G(\psi^j)\right],
\end{equation}
where $\zeta_N=e^{2\pi\sqrt{-1}/N}$ and $\psi$ is a primitive multiplicative character of order $N$ over $\gf(r)^*$.

From (\ref{equ-G sum& G period}), one knows that the values of the Gaussian periods in general are also very hard to compute.
However, they can be computed in a few cases. To present some known
results on Gaussian periods, we need to introduce period polynomials.

The {\em period polynomials} $\psi_{(N,r)}(X)$ are defined by
$$
\psi_{(N,r)}(X)=\prod_{i=0}^{N-1} \left(X - \eta_i^{(N,r)} \right).
$$
It is known that $\psi_{(N,r)}(X)$ is a polynomial with integer coefficients
\cite{Myer}. We will need the following four lemmas whose proofs can be found
in \cite{Myer}.

\begin{lemma}\label{lem-period1}
Let $N=3$. Let $c$ and $d$ be defined by $4r=c^2+27d^2$, $c \equiv 1 \pmod{3}$, and, if
$p \equiv 1 \pmod{3}$, then $\gcd(c,p)=1$. These restrictions determine $c$ uniquely,
and $d$ up to sign. Then we have
\begin{eqnarray*}\label{eqn-period11}
\psi_{(3,r)}(X)= X^3 + X^2 - \frac{r-1}{3} X - \frac{(c+3)r-1}{27}.
\end{eqnarray*}
\end{lemma}

\vspace{.2cm}
\begin{lemma}\label{lem-period2}
Let $N=3$.
We have the following results on the factorization of $\psi_{(3,r)}(X)$.
\begin{itemize}
\item[(a)] If $p \equiv 2 \pmod{3}$, then $ms$ is even, and
\begin{eqnarray*}
\psi_{(3,r)}(X) =
 \left\{ \begin{array}{ll}
                         3^{-3}(3X+1+2\sqrt{r})(3X+1-\sqrt{r})^2 & \mbox{ if $sm/2$ even,} \\
                         3^{-3}(3X+1-2\sqrt{r})(3X+1+\sqrt{r})^2 & \mbox{ if $sm/2$ odd.}
                        \end{array}
                \right.
\end{eqnarray*}
\item[(b)] If $p \equiv 1 \pmod{3}$, and $sm \not\equiv 0 \pmod{3}$, then $\psi_{(3,r)}(X)$
is irreducible over the rationals.
\item[(c)] If $p \equiv 1 \pmod{3}$, and $sm \equiv 0 \pmod{3}$, then
\begin{eqnarray*}
\psi_{(3,r)}(X) = \frac{1}{27}(3X+1-c_1r^{\frac{1}{3}})
                   \left(3X+1+\frac{1}{2}(c_1+9d_1)r^{\frac{1}{3}}\right)
                   \left(3X+1+\frac{1}{2}(c_1-9d_1)r^{\frac{1}{3}}\right),
\end{eqnarray*}
where $c_1$ and $d_1$ are given by $4p^{sm/3}=c_1^2+27d_1^2$, $c_1 \equiv 1 \pmod{3}$ and
$\gcd(c_1,p)=1$.
\end{itemize}
\end{lemma}

\vspace{.2cm}
\begin{lemma}\label{lem-period3}
Let $N=4$. Let $u$ and $v$ be defined by $r=u^2+4v^2$, $u \equiv 1 \pmod{4}$, and, if
$p \equiv 1 \pmod{4}$, then $\gcd(u,p)=1$. These restrictions determine $u$ uniquely,
and $v$ up to sign.

If $n$ is even, then
\begin{eqnarray*}\label{eqn-period21}
\psi_{(4,r)}(X) =
 X^4+X^3 - \frac{3r-3}{8}X^2 +
                    \frac{(2u-3)r+1}{16}X +
                   \frac{r^2-(4u^2-8u+6)r+1 }{256}.
\end{eqnarray*}

If $n$ is odd, then
\begin{eqnarray*}\label{eqn-period22}
\psi_{(4,r)}(X)=
 X^4+X^3 + \frac{r+3}{8}X^2 + \frac{(2u+1)r+1}{16}X
                 + \frac{9r^2-(4u^2-8u-2)r+1}{256}.
\end{eqnarray*}
\end{lemma}

\vspace{.2cm}
\begin{lemma}\label{lem-period4}
Let $N=4$.
We have the following results on the factorization of $\psi_{(4,r)}(X)$.
\begin{itemize}
\item[(a)] If $p \equiv 3 \pmod{4}$, then $ms$ is even, and
\begin{eqnarray*}
\psi_{(4,r)}(X)=
 \left\{ \begin{array}{ll}
                         4^{-4}(4X+1+3\sqrt{r})(4X+1-\sqrt{r})^3 & \mbox{ if $sm/2$ even,} \\
                         4^{-4}(4X+1-3\sqrt{r})(4X+1+\sqrt{r})^3 & \mbox{ if $sm/2$ odd.}
                        \end{array}
                \right.
\end{eqnarray*}
\item[(b)] If $p \equiv 1 \pmod{4}$, and $sm$ is odd, then $\psi_{(4,r)}(X)$
is irreducible over the rationals.
\item[(c)] If $p \equiv 1 \pmod{4}$, and $sm \equiv 2 \pmod{4}$, then
\begin{eqnarray*}
\psi_{(4,r)}(X) =
  4^{-4} \left( (4X+1)^2 + 2\sqrt{r} (4X+1) -r -2 \sqrt{r}u \right) \times  \\
                      \left( (4X+1)^2 - 2\sqrt{r} (4X+1) -r +2 \sqrt{r}u \right),
\end{eqnarray*}
the quadratics being irreducible, the $u$ is defined in Lemma \ref{lem-period3}.

\item[(d)] If $p \equiv 1 \pmod{4}$, and $sm \equiv 0 \pmod{4}$, then
\begin{eqnarray*}
\begin{array}{r}
\psi_{(4,r)}(X) =
 4^{-4} \left( (4X+1) + \sqrt{r} + 2 r^{1/4}u_1 \right) \left( (4X+1) + \sqrt{r} - 2 r^{1/4}u_1 \right) \\
                   \times    \left( (4X+1) - \sqrt{r} + 4 r^{1/4}v_1 \right) \left( (4X+1) - \sqrt{r} - 4 r^{1/4}v_1 \right)
\end{array}
\end{eqnarray*}
where $u_1$ and $v_1$ are given by $p^{sm/2}=u_1^2+4v_1^2$, $u_1 \equiv 1 \pmod{4}$ and
$\gcd(u_1,p)=1$.
\end{itemize}
\end{lemma}

\vspace{.2cm}
The following lemma follows from Lemma \ref{lem-quadratic} and (\ref{eqn-etachi1}).

\vspace{.2cm}
\begin{lemma}\label{lem-degree2}
When $N=2$, the Gaussian periods are given by the following:
\begin{eqnarray*}
\eta_0^{(2,r)}=
\left\{
\begin{array}{ll}
\frac{-1+(-1)^{sm-1}r^{1/2}}{2} & \mbox{if $p\equiv 1 \pmod{4}$} \\
\frac{-1+(-1)^{sm-1}(\sqrt{-1})^{sm} r^{1/2}}{2} & \mbox{if $p\equiv 3 \pmod{4}$}
\end{array}
\right.
\end{eqnarray*}
and
$$
\eta_1^{(2,r)} = -1 - \eta_0^{(2,r)}.
$$
\end{lemma}

By Lemma \ref{lemma2-semi} and (\ref{equ-G sum& G period}), the Gaussian periods in the semi-primitive case are known and are described in
the following lemma \cite{BM72,Myer} .

\vspace{.2cm}
\begin{lemma}\label{lem-semip}
Assume that $N>2$ and there exists a positive integer $j$ such that $p^j \equiv -1 \pmod{N}$, and
the $j$ is the least such. Let $r=p^{2j\gamma}$ for some integer $\gamma$.

(a) If $\gamma$, $p$ and $(p^j+1)/N$ are all odd, then
\begin{eqnarray*}
\begin{array}{l}
\eta_{N/2}^{(N,r)}=\frac{(N-1)\sqrt{r}-1}{N}, \\
\eta_{k}^{(N,r)}=-\frac{\sqrt{r}+1}{N}  \mbox{ for } k \ne N/2.
\end{array}
\end{eqnarray*}

(b) In all other cases,
\begin{eqnarray*}
\begin{array}{l}
\eta_{0}^{(N,r)}=\frac{(-1)^{\gamma+1}(N-1)\sqrt{r}-1}{N}, \\
\eta_{k}^{(N,r)}=\frac{(-1)^\gamma \sqrt{r}-1}{N} \mbox{ for } k \ne 0.
\end{array}
\end{eqnarray*}
\end{lemma}

From Lemma \ref{lemma-index2} and (\ref{equ-G sum& G period}), the Gaussian periods in the so-called quadratic residues (or index 2) case can be also computed.
The results with $3\neq N\equiv 3\pmod{4}$ being odd prime are given by \cite{BM73,Myer}.


\section{The weights in irreducible cyclic codes}

Let $N>1$ be an integer dividing $r-1$, and put $n=(r-1)/N$.
Let $\alpha$ be a primitive element of $\gf(r)$ and let $\theta=\alpha^N$.
Let $Z(r,a)$ denote the number of solutions $x \in \gf(r)$ of the equation $\tr_{r/q}(ax^{N})=0$.
Let $\zeta_p=e^{2\pi \sqrt{-1}/p}$, and $\chi(x)=\zeta_p^{\tr_{r/p}(x)}$, where $\tr_{r/p}$ is
the trace function from $\gf(r)$ to $\gf(p)$. Then $\chi$ is an
additive character of $\gf(r)$. We have then by Lemma \ref{lem-haw}
\begin{eqnarray}\label{eqn-forall1}
 Z(r,a)
&=& \frac{1}{q} \sum_{y \in \gf(q)} \sum_{x \in \gf(r)} \zeta_p^{\tr_{q/p}(y \tr_{r/q} (ax^{N}))} \nonumber \\
&=& \frac{1}{q} \sum_{y \in \gf(q)} \sum_{x \in \gf(r)} \chi(yax^{N}) \nonumber \\
&=& \frac{1}{q} \left[ q + r-1 + \sum_{y \in \gf(q)^*} \sum_{x \in \gf(r)^*} \chi(yax^{N}) \right] \nonumber \\
& =& \frac{1}{q} \left[ q + r-1 + N \sum_{y \in \gf(q)^*} \sum_{x \in C_{0}^{(N,r)}} \chi(ya x) \right] \nonumber \\
& =& \frac{1}{q} \left[ q + r-1 + (q-1)\gcd((r-1)/(q-1), N) \sum_{z \in C_{0}^{\left(\gcd\left(\frac{r-1}{q-1},N\right),r\right)}} \chi(az) \right]
\end{eqnarray}

Then the Hamming weight of the codeword
$$
(\tr_{r/q}(\beta), \tr_{r/q}(\beta \theta), ...,
              \tr_{r/q}(\beta \theta^{n-1}))
$$
in the irreducible cyclic code of (\ref{eqn-irrecode}) is equal to
\begin{eqnarray}\label{eqn-wtmain}
n-\frac{Z(r,\beta)-1}{N}=
\frac{(q-1)\left(r-1-\gcd\left(\frac{r-1}{q-1}, N\right) \eta_{k}^{ \left(\gcd\left(\frac{r-1}{q-1},N\right),r\right)} \right)}{qN}.
\end{eqnarray}

The weight expression of (\ref{eqn-wtmain}) is the key observation of this paper and
proves that the determination of the weight distribution of an irreducible cyclic code
is equivalent to that of the Gaussian periods of order $N_1=\gcd((r-1)/(q-1), N)$. McEliece \cite{McEl74} gave a different proof of (\ref{eqn-wtmain}) by Gaussian sums, and from (\ref{equ-G sum& G period}), we know that the weights of an irreducible cyclic
code can be expressed as a linear combination of Gaussian sums.

\begin{theorem} \label{thm-wdivisibility}
Let $N_1=\gcd((r-1)/(q-1), N)$. Then, for all $i$ with $0 \le i \le N_1-1$, we have
\par (i)~$\eta_i^{(N_1,r)}\in\mathbb{Z}$;
\par (ii)~$N_1 \eta_i^{(N_1,r)} + 1 \equiv 0 \pmod{q}$; and
\par (iii)~$\left|\eta_i^{(N_1,r)}+\frac{1}{N_1}\right|\leqslant \left\lfloor{\frac{(N_1-1)\sqrt{r}}{N_1}} \right\rfloor$.
\end{theorem}

\begin{proof}
The conclusions of Parts (i) and (ii) follow from  (\ref{eqn-wtmain}) directly, and  that of Part (iii) follows from (\ref{equ-Gvalue}) and (\ref{equ-G sum& G period}).
\end{proof}

Theorem \ref{thm-wdivisibility} is an interesting result in the theory of cyclotomy.

\begin{theorem} \label{thm-gaussianp}
Let $N_1=\gcd((r-1)/(q-1), N)$. Then the Hamming weight of every codeword
in the irreducible cyclic code of (\ref{eqn-irrecode}) is divisible by
$$
\frac{(q-1)}{\gcd\left(q-1,  N/N_1 \right)}.
$$
\end{theorem}

\begin{proof}
By (\ref{eqn-wtmain}), the Hamming weight of every nonzero codeword is equal to
$$
\frac{q-1}{\gcd(q-1, N/N_1)} \frac{r-(1+N_1\eta_k)}{q \frac{N}{\gcd(q-1, N/N_1)}}.
$$
The desired conclusion then follows from the fact that
$$
\gcd\left(q-1, q \frac{N}{\gcd(q-1, N/N_1)}\right)=1.
$$

\end{proof}

Particularly, when $N$ divides $(r-1)/(q-1)$, the Hamming weight of every codeword
in the irreducible cyclic code of (\ref{eqn-irrecode}) is divisible by $q-1$.

\begin{example}
Let $q=5$. $m=4$, $N=4$. Then the  irreducible cyclic code of (\ref{eqn-irrecode}) over
$\gf(q)$ has length , dimension, and the following weight distribution:
$$
1+156x^{112} + 156 x^{124} + 156 x^{128} + 156 x^{136}.
$$
So by Theorem \ref{thm-gaussianp},
4 is a common divisor of all nonzero weights. Note that
$$
\gcd(112, 124, 128, 136)=4.
$$
\end{example}

\begin{example}
Let $q=3$. $m=4$, $N=2$. Then the  irreducible cyclic code of (\ref{eqn-irrecode}) over
$\gf(q)$ has length 40, dimension 4, and the following weight distribution:
$$
1+ 40x^{24} + 40 x^{30}.
$$
So by Theorem \ref{thm-gaussianp},
2 is a common divisor of all nonzero weights. Note that
$
\gcd(24, 30)=6.
$
\end{example}

\section{The weight distribution in the case that $\gcd((r-1)/(q-1), N)=1$}

\begin{theorem}\label{thm-2wt0}
Let $N$ be a positive divisor of $r-1$ such that  $\gcd((r-1)/(q-1), N)=1$.
Then the set
$\C(r,N)$ in (\ref{eqn-irrecode}) is a $[(q^m-1)/N, m, (q-1)q^{m-1}/N]$ constant-weight code
with the weight enumerator
$$
1 + (r-1) x^{\frac{(q-1)q^{m-1}}{N}  }.
$$
\end{theorem}

\begin{proof}
Since $N$ divides $r-1$ and $\gcd((r-1)/(q-1), N)=1$, $N$ must divide $q-1$.
It follows that
$$
\gcd((r-1)/(q-1), N)=\gcd(m,N)=1.
$$

Let $\alpha$ be the generator of $\gf(r)^*$. For any $a \ne 0$, it
follows from (\ref{eqn-wtmain}) and Lemma \ref{lem-degree2} that  for any $\beta \in \gf(r)^*$ the Hamming weight of any codeword
$$
\bc(\beta)=(\tr_{r/q}(\beta), \tr_{r/q}(\beta \theta), ..., \tr_{r/q}(\beta \theta^{n-1})
$$
of the code $\C(r,N)$ is equal to
$$
n-\frac{Z(r,\beta)-1}{N}=\frac{(q-1)q^{m-1}}{N}.
$$
Note that $|C_0^{(2,r)}|=|C_1^{(2,r)}|=(r-1)/2$. The weight distribution and dimension of the code follow.
This completes the proof.
\end{proof}

\begin{theorem}\label{thm-1wt2}
Let $N$ be a positive divisor of $r-1$.
Then the set
$\C(r,N)$ in (\ref{eqn-irrecode}) is a $[(q^m-1)/N, m]$ constant-weight code  if
and only if $\gcd((r-1)/(q-1), N)=1$.
\end{theorem}

\begin{proof}
Theorem \ref{thm-2wt0} shows that the condition is sufficient. We now prove the necessity of
the condition.  Let $N_1=\gcd((r-1)/(q-1), N)$ and $n_1=(r-1)/N_1$. Assume that $\C(r,N)$ is a constant weight code.
It then follows from (\ref{eqn-wtmain}) that $1+N_1\eta_i$ is a constant $\lambda$ for all $i$. Define
$\zeta_i=1+N_1\eta_i$. Then the formulas in  Lemma \ref{lem-periodpro} becomes
\begin{enumerate}
\item $\sum_{i=0}^{N_1-1} \zeta_i = 0.$
\item $\sum_{i=0}^{N_1-1} \zeta_i \zeta_{i+k} = N_1(N_1\theta_k-1)r$ for all $k \in \{0,1, \cdots, N_1-1\}$, where
$$
\theta_k=\left\{ \begin{array}{ll}
                          1 & \mbox{ if $n_1$ is even and $k=0$} \\
                          1 & \mbox{ if $n_1$ is odd and $k=N_1/2$} \\
                          0 & \mbox{ otherwise,}
                          \end{array}
                          \right.
$$
\end{enumerate}
and equivalently $\theta_k=1$ if and only if $-1 \in C_{k}^{(N_1,r)}$.

Since $N_1$ is a divisor of $(r-1)/(q-1)$, $\gf(q)^* \subset C_0^{(N_1, r)}$.  It follows that $\theta_0=1$. Hence,
we have
$$
N_1\lambda = 0, \ N_1 \lambda^2 = N_1(N_1-1)r.
$$
Whence, $N_1=1$.  This completes the proof.
\end{proof}

Theorem  \ref{thm-1wt2}  above is a complete characterization of one-weight irreducible cyclic codes in the
general case that $N$ is any divisor of $r-1$, which is different from Theorem 1 in \cite{Vega}, where  Vega
and Wolfmann considered only the case that $N$ is a divisor of $q-1$ and use the period of the check polynomial
of the code for the characterization.  Theorem \ref{thm-2wt0}  is extension of Theorem 6 in \cite{DingIT09}.

\section{The weight distribution in the case that $\gcd((r-1)/(q-1), N)=2$}

\begin{theorem}\label{thm-2wt1}
Let $N$ be a positive divisor of $r-1$.
If $\gcd((r-1)/(q-1), N)=2$, then the set
$\C(r,N)$ in (\ref{eqn-irrecode}) is a $[(q^m-1)/N, m, (q-1)(r - \sqrt{r})/Nq]$ two-weight code
with the weight enumerator
$$
1 + \frac{r-1}{2} x^{\frac{(q-1)(r - \sqrt{r}) }{qN} } + \frac{r-1}{2} x^{\frac{(q-1)(r + \sqrt{r})}{qN}}.
$$
\end{theorem}

\begin{proof}
Since $\gcd((r-1)/(q-1), N)=2$, $m$ is even and $q$ is odd.
Let $\alpha$ be the generator of $\gf(r)^*$.
Let $a \in C_{h}^{(2,r)}$. It then
follows from (\ref{eqn-wtmain}) and Lemma \ref{lem-degree2} that for any $\beta \in \gf(r)^*$ the Hamming weight of any codeword
$$
\bc(\beta)=(\tr_{r/q}(\beta), \tr_{r/q}(\beta \theta), ..., \tr_{r/q}(\beta \theta^{n-1})
$$
of the code $\C(r,N)$ is equal to
$$
n-\frac{Z(r,\beta)-1}{N}=\frac{(q-1)(r \mp \sqrt{r})}{qN}>0.
$$
Note that $|C_0^{(2,r)}|=|C_1^{(2,r)}|=(r-1)/2$. The weight distribution and dimension of the code follow.
This completes the proof.
\end{proof}

Theorem \ref{thm-2wt1} is an extension of Theorem 7 in Baumert and McEliece \cite{BM72}.

\vspace{.2cm}
\begin{example}
Let $q=9$, $m=2$, and $N=q-1=8$. Then $\gcd((r-1)/(q-1), N)=2$. All the
conditions of Theorem \ref{thm-2wt1} are satisfied. The set $\C(r,8)$ is then
a $[10, 2, 8]$ code over $\gf(9)$
with the weight distribution $1+40x^{8} + 40x^{10}$.
\end{example}

\begin{example}
Let $q=9$, $m=2$, and $N=2(q-1)=16$. Then $\gcd((r-1)/(q-1), N)=2$. All the
conditions of Theorem \ref{thm-2wt1} are satisfied. The set $\C(r,16)$ is then
a $[5, 2, 4]$ code over $\gf(9)$
with the weight distribution $1+40x^{4} + 40x^{5}$.
\end{example}

\begin{example}
Let $q=3$, $m=4$, and $N=q-1=2$. Then $\gcd((r-1)/(q-1), N)=2$. All the
conditions of Theorem \ref{thm-2wt1} are satisfied. The set $\C(r,2)$ is then
a $[40, 4, 24]$ code over $\gf(3)$
with the weight distribution $1+40x^{24} + 40x^{30}$.
\end{example}

\begin{example}
Let $q=3$, $m=4$, and $N=2(q-1)=4$. Then $\gcd((r-1)/(q-1), N)=4$. The set $\C(r,4)$ is then
a $[20, 4, 12]$ code over $\gf(3)$
with the weight distribution $1+60x^{12} + 20x^{18}$.  In this case, the weight distribution
of this code is different from the one in Theorem \ref{thm-2wt1}.
\end{example}

\section{The weight distribution in the case that $\gcd((r-1)/(q-1), N)=3$}

\begin{theorem}\label{thm-2wt1111}
Let $N$ be a divisor of $r-1$.
When $\gcd((r-1)/(q-1), N)=3$ and $p \equiv 1 \pmod{3}$, the set
$\C(r,N)$ in (\ref{eqn-irrecode}) is a $[(q^m-1)/N), m]$ code with the
following weight distribution:
\begin{eqnarray*}
1 + \frac{r-1}{3}x^{\frac{(q-1)(r-c_1r^{1/3})}{Nq}} +
    \frac{r-1}{3}x^{\frac{(q-1)[r+\frac{1}{2}(c_1+9d_1)r^{1/3}]}{Nq}} +
    \frac{r-1}{3}x^{\frac{(q-1)[r+\frac{1}{2}(c_1-9d_1)r^{1/3}]}{Nq}},
\end{eqnarray*}
where $c_1$ and
$d_1$ are uniquely given by $4q^{m/3}=c_1^2+27d_1^2$, $c_1 \equiv 1 \pmod{3}$ and
$\gcd(c_1,p)=1$.

\end{theorem}

\vspace{.2cm}
\begin{proof}
By assumption $\gcd(m, q-1)=3$. It then follows from (\ref{eqn-fgfg0}) that
\begin{eqnarray*}
\left\{xy: y \in \gf(q)^*, \ x \in C_i^{(N,r)}\right\} = \frac{3(q-1)}{N}*C_i^{(3,r)}.
\end{eqnarray*}

Since $\gcd((r-1)/(q-1), N)=3$, $(r-1)/(q-1) \bmod{3} = m \bmod{3} = 0$.
Note that every element of $\gf(q)^*$ is of the form
$\alpha^{i(r-1)/(q-1)}$ for some integer $i$. Hence, $\gf(q)^* \subset C_0^{(3,r)}$.
It then follows from Lemma \ref{lem-period2} that the Gaussian periods
$\eta_i^{(3,r)}$ take only the following three distinct values:
$$
\frac{-1+c_1r^{1/3}}{3},  \frac{-1-\frac{1}{2}(c_1+9d_1)r^{1/3}}{3},  \frac{-1-\frac{1}{2}(c_1-9d_1)r^{1/3}}{3}.
$$

It then follows from (\ref{eqn-wtmain}) that  for any $\beta \in \gf(r)^*$ the Hamming weight of any codeword
$$
\bc(\beta)=(\tr_{r/q}(\beta), \tr_{r/q}(\beta \theta), ..., \tr_{r/q}(\beta \theta^{n-1})
$$
of the code $\C(r,q-1)$ is equal to
$$
n-\frac{Z(r,\beta)-1}{N}=\frac{1}{q} \left[ q + r-1 + 3 (q-1) \eta_i^{(3,r)} \right]>0.
$$
Note that $|C_i^{(3,r)}|=(r-1)/3$. The weight distribution and dimension of the code then follow.
This completes the proof.
\end{proof}

Theorem  \ref{thm-2wt1111}  of this section is an extension of Theorem 14 in \cite{DingIT09} and Theorem 6 in \cite{Ding09} .

\vspace{.2cm}
\begin{example}
Let $q=7$, $m=3$ and $N=q-1=6$. Then the set $\C(r,N)$ in (\ref{eqn-irrecode}) is a $[57, 3, 45]$ code with the
weight distribution $1+114x^{45} + 114 x^{48} + 114 x^{54}$.
\end{example}

\begin{example}
Let $q=7$, $m=3$ and $N=3(q-1)=18$. Then the set $\C(r,N)$ in (\ref{eqn-irrecode}) is a $[19, 3, 15]$ code with the
weight distribution $1+114x^{15} + 114 x^{16} + 114 x^{27}$.
\end{example}

\begin{theorem}\label{thm-e137}
Let $N$ be a divisor of $r-1$.
Suppose that $\gcd((r-1)/(q-1), N)=3$ and $p \equiv 2 \pmod{3}$.
If $sm \equiv 0 \pmod{4}$, then
$\C(r,N)$ is  a $[(r-1)/N, m, (q-1)(r-\sqrt{r})/Nq]$ code over $\gf(q)$
with the weight distribution
$$
1 + \frac{2(r-1)}{3} x^{\frac{(q-1)(r-\sqrt{r})}{Nq}} + \frac{r-1}{3} x^{\frac{(q-1)(r+2\sqrt{r})}{Nq}}.
$$

If $sm \equiv 2 \pmod{4}$, then
$\C(r,N)$ is a $[(r-1)/N, m, (q-1)(r-2\sqrt{r})/Nq]$ code over $\gf(q)$
with the weight distribution
$$
1 + \frac{r-1}{3} x^{\frac{(q-1)(r-2\sqrt{r})}{Nq}} + \frac{2(r-1)}{3} x^{\frac{(q-1)(r+\sqrt{r})}{Nq}}.
$$
\end{theorem}

\vspace{.2cm}
\begin{proof}
Note that $\gcd((r-1)/(q-1), N)=3$ and $p \equiv 2 \pmod{3}$.  This theorem becomes a
special case of Theorem \ref{thm-semipri}.
\end{proof}

\vspace{.2cm}
\begin{example}
Let $q=4$, $m=6$ and $N=q-1=3$. Then the set $\C(r,N)$ in (\ref{eqn-irrecode}) is a $[1365, 6, 1008]$ code
over $\gf(4)$ with the
weight distribution $1+ 2730 x^{1008} + 1365 x^{1056}$.
\end{example}

\vspace{.2cm}
\begin{example}
Let $q=4$, $m=6$ and $N=3(q-1)=9$. Then the set $\C(r,N)$ in (\ref{eqn-irrecode}) is a $[455, 6, 336]$ code
over $\gf(4)$ with the
weight distribution $1+ 2730 x^{336} + 1365 x^{352}$.
\end{example}

\vspace{.2cm}
\begin{example}
Let $q=4$, $m=3$ and $N=q-1=3$. Then the set $\C(r,N)$ in (\ref{eqn-irrecode}) is a $[21, 3, 12]$ code
over $\gf(4)$ with the
weight distribution $1+ 21x^{12} + 42 x^{18}$.
\end{example}

\vspace{.2cm}
\begin{example}
Let $q=4$, $m=3$ and $N=3(q-1)=9$. Then the set $\C(r,N)$ in (\ref{eqn-irrecode}) is a $[7, 3, 4]$ code
over $\gf(4)$ with the
weight distribution $1+ 21x^{4} + 42 x^{6}$.
\end{example}

\section{The weight distribution in the case that $\gcd((r-1)/(q-1), N)=4$}

\begin{theorem}\label{thm-ccc4}
Let $N$ be a divisor of $r-1$.
If $\gcd((r-1)/(q-1), N)=4$ and $p \equiv 1 \pmod{4}$, $\C(r,N)$ is  a $[(r-1)/N, m]$ code over $\gf(q)$
with the weight distribution
\begin{eqnarray*}
1+ \frac{r-1}{4}x^{\frac{(q-1)(r+\sqrt{r}+2u_1r^{1/4})}{Nq}} + \frac{r-1}{4}x^{\frac{(q-1)(r+\sqrt{r}-2u_1r^{1/4})}{Nq}}  \\
 + \frac{r-1}{4}x^{\frac{(q-1)(r-\sqrt{r}+4v_1r^{1/4})}{Nq}} + \frac{r-1}{4}x^{\frac{(q-1)(r-\sqrt{r}-4v_1r^{1/4})}{Nq}}
\end{eqnarray*}
where $u_1$ and $v_1$ are given by $q^{m/2}=u_1^2+4v_1^2$, $u_1 \equiv 1 \pmod{4}$, and
$\gcd(u_1,p)=1$.

If  $\gcd((r-1)/(q-1), N)=4$ and $p \equiv 3 \pmod{4}$, $\C(r,N)$ is  a $[(r-1)/N, m]$ code over $\gf(q)$
with the weight distribution
\begin{eqnarray*}
1+ \frac{3(r-1)}{4} x^{\frac{(q-1)(r-\sqrt{r})}{Nq}} + \frac{r-1}{4} x^{\frac{(q-1)(r+3\sqrt{r})}{Nq}} .
\end{eqnarray*}
\end{theorem}

\vspace{.2cm}
\begin{proof}
Note that $\gcd((r-1)/(q-1), N)=4$. Then similar to the proof of Theorem \ref{thm-2wt1111},
we can prove the weight distribution formula with the help of Lemma \ref{lem-period4}
and (\ref{eqn-wtmain}).
\end{proof}

Theorem  \ref{thm-ccc4}  of this section is an extension of Theorem 21 in \cite{DingIT09} and Theorem 7 in \cite{Ding09}.

\begin{example}
Let $q=5$, $m=4$ and $N=q-1=4$. Then the set $\C(r,N)$ in (\ref{eqn-irrecode}) is a $[156, 4, 112]$ code over $\gf(5)$
with the weight distribution $1+  156x^{112} + 156x^{124} + 156x^{128} + 156 x^{136}$.
\end{example}

\begin{example}
Let $q=5$, $m=4$ and $N=4(q-1)=16$. Then the set $\C(r,q-1)$ in (\ref{eqn-irrecode}) is a $[39, 4, 28]$ code over $\gf(5)$
with the weight distribution $1+  156x^{28} + 156x^{31} + 156x^{32} + 156 x^{34}$.
\end{example}

\section{The weight distribution in the quadratic residue case}

In another special case, called the ``quadratic residue" or ``index 2" case, the weight distribution of the irreducible cyclic code is
known and described in the following theorem.

\begin{theorem}\label{thm-quadra}
Let notations be defined as in Lemma \ref{lemma-index2}. For $0\leqslant i\leqslant N_1-1$, define
$$\left\{\begin{array}{l}
  i_2:=v_l(i),\ \mbox{i.e., }l^{i_2}\parallel i;\\
  i_1:=i/l^{i_2}\in(\mathbb{Z}/l^{\lambda-i_2}\mathbb{Z})^*.
\end{array}\right.$$
Then, the Hamming weight of the codeword $c(\beta)$ with $\beta\in C_i^{(r,N_1)}$ is given by
$$\begin{array}{rl}
w_H(c(\beta))&=\dfrac{(q-1)}{Nq}\left[r-\sum\limits_{u=1}^{l^\lambda-1}G(\psi^u,\chi_1)\psi^{-u}(g^i)\right]\\
    &=\dfrac{(q-1)}{Nq}\left[r-\sum\limits_{t=0}^{i_2}l^t\left(A_t^{(s,\lambda)}P_t^{(s,\lambda)}-A_{t+1}^{(s,\lambda)}P_{t+1}^{(s,\lambda)}\right)-\left(\frac{i_1}{l}\right)l^{i_2+1}P_{i_2+1}^{(s,\lambda)}B_{i_2+1}^{(s,\lambda)}\right],
\end{array}$$
where we take $A_0=A_{\lambda+1}=B_{\lambda+1}=0$.
\end{theorem}

\begin{proof}
The conclusions of this theorem follow from (\ref{eqn-wtmain}), Lemma \ref{lemma-index2} and the
conditions stated in this theorem.
\end{proof}

Regarding Theorem \ref{thm-quadra}, we have the following remarks.
\begin{itemize}
\item Theorem \ref{thm-quadra} is an extension of the main results obtained by Baumert and
Mykkeltveit \cite{BM73} and the main results of \cite[\S11.7]{BEW}.
\item With the explicit formulas of Theorem \ref{thm-quadra} and the recursive relation of $A_t^{(s,\lambda)},~B_t^{(s,\lambda)},~P_t^{(s,\lambda)}$ with respect to $\lambda$, one can derive the recursive algorithms presented in \cite{Moisio}.
\item According to the conclusions of \cite{Y-X10}, there are six subcases for Gauss sums in the index 2 case. Theorem \ref{thm-quadra} is the corresponding result for one of the six subcases.
\end{itemize}

\begin{example}
Let $q=2$, $m=42$ and $N=7^2=49$. Then the set $\C(r,N)$ in (\ref{eqn-irrecode}) is a $[89756051247, 42, 44877307904]$ code over $\gf(2)$
with the weight distribution $$1+ x^{44877307904} + 3x^{44877832192} + 21x^{44877979648} + 21x^{44878086144} +3x^{44878356480}.$$
\end{example}

\begin{example}
Let $q=3$, $m=55$ and $N=11^2=121$. Then the set $\C(r,N)$ in (\ref{eqn-irrecode}) is a $$[1441729016604299000588186, 55, 961152677733830625644778]$$ code over $\gf(3)$
with the weight distribution $$1+ 6x^{961152677733830625644778} + 55x^{961152677735964537698190} + 55x^{961152677736445713945528} + 5x^{961152677738914357301436}.$$
\end{example}

\section{The weight distribution in the case that $n$ is prime power}

The following result is presented in \cite{India}.

\begin{theorem}\label{thm-nwt0}
Let $q=p^s$. Let $t$ be an odd prime and $\ell$ be a positive integer. Assume
that the multiplicative order of $q$ mudulo $t^\ell$ is $t^{d}$, where $0 \le d <\ell$.
Define $m=t^d$ and $N=(q^m-1)/t^j$ for any $j$ with $1 \le j \le \ell$.

If $j \le \ell - d$, then the set
$\C(r,N)$ in (\ref{eqn-irrecode}) is a $[t^j, 1, t^j]$ constant-weight code over $\gf(q)$
with the weight enumerator
$$
1 + (q-1) x^{t^j}.
$$

If $j > \ell - d$, then the set
$\C(r,N)$ in (\ref{eqn-irrecode}) is a $[t^j, t^{j-(\ell -d)}]$ cyclic code over $\gf(q)$
with the weight enumerator
$$
\sum_{w=0}^{t^{r-\ell +d}}   {t^{j-\ell+d} \choose w} x^{t^{(\ell-d)}w}.
$$

\end{theorem}

\begin{example}
Let $q=2^2$ and $t^\ell=3^3$. Then the order of $q$ modulo $t^\ell$ is $3^2$.
Define $m=3^2=9$ and $N=(q^m-1)/t^2$. Then $n=t^2=9$, and the set
$\C(r,N)$ in (\ref{eqn-irrecode}) is a $[9, 3, 3]$ cyclic code over $\gf(4)$
with the weight enumerator
$$
1+9x^3+27x^6+27x^9.
$$
\end{example}

\section{The weight distribution in the semi-primitive and related cases}

\begin{theorem}\label{thm-semipri}
Let $p$ be a prime and $sm$ be even.
Let $N$ be a positive divisor of $r-1$ and $N_1=\gcd((r-1)/(q-1), N)>2$.
Assume there exists a positive integer $j$ such that $p^j \equiv -1 \pmod{N_1}$, and
the $j$ is the least such. Define $\gamma=sm/2j$.

(a) If $\gamma$, $p$ and $(p^j+1)/N_1$ are all odd, then the set
$\C(r,N)$ in (\ref{eqn-irrecode}) is a $[(q^m-1)/N, m]$ code  over $\gf(q)$
with the weight enumerator
$$
1 + \frac{r-1}{N_1}x^{\frac{(q-1)(r-(N_1-1)\sqrt{r})}{qN}}  + \frac{(r-1)(N_1-1)}{N_1}x^{\frac{(q-1)(r+\sqrt{r})}{qN}} ,
$$
provided that $N_1<\sqrt{r}+1$.

(b) In all other cases, the set
$\C(r,N)$ in (\ref{eqn-irrecode}) is a $[(q^m-1)/N, m]$  code
with the weight enumerator
$$
1 + \frac{r-1}{N_1}x^{\frac{(q-1)(r+(-1)^\gamma (N_1-1)\sqrt{r})}{qN}}  + \frac{(r-1)(N_1-1)}{N_1}x^{\frac{(q-1)(r-(-1)^\gamma\sqrt{r})}{qN}} ,
$$
provided that $\sqrt{r} + (-1)^\gamma (N_1-1)>0$.
\end{theorem}

\begin{proof}
The conclusions of this theorem follow from (\ref{eqn-wtmain}), Lemma \ref{lem-semip} and the
conditions stated in this theorem.
\end{proof}

Regarding Theorem \ref{thm-semipri}, we have the following remarks.
\begin{itemize}
\item When $N_1=N$, this is the classical semi-primitive case, and the weight distribution
          of the code was studied by Delsarte and Goethals \cite{DG70}, McEliece \cite{McEl},
      and Baumert and McEliece \cite{BM72}.
\item When $N_1<N$, this may not be the semiprimitive case for $N$.
        For example, let $q=7$, $m=2$ and $N=12$. We now prove that this is not the
        semi-primitive case for $N=12$. To this end, we prove that there is no positive integer $j$ such
        that $7^j \equiv -1 \pmod{12}$, which is equivalent to the following system of
       congruences:
       $$
       7^j \equiv -1 \pmod{4} \mbox{ and } 7^j \equiv -1 \pmod{3}
       $$
       by the Chinese Remainder Theorem. The second congruence does not have a solution.

     In this case $N_1=4|7^1+1$. By Theorem \ref{thm-semipri} the code over $\gf(7)$ has
     length $4$, dimension $2$ and weight enumerator
    $$
      1+12x^{2}+36x^{4}.
    $$
   This shows that some non-semiprimitive cases can be settled using the results of the
   semiprimitive cases.
\item The condition that $N_1<\sqrt{r}+1$ or  $\sqrt{r} + (-1)^\gamma (N_1-1)>0$ is
          to ensure that the dimension of the code is $m$.
\item Theorem 2.1 in \cite{DLN} is a special case of Theorem \ref{thm-semipri} above.
\end{itemize}

Theorem \ref{thm-semipri} describes a class of two-weight irredicuble cyclic codes
over $\gf(q)$, and is an extension of Theorem 6 in Baumert and McEliece \cite{BM72}.
It is an interesting problem to find out all two-weight irreducible cyclic
codes over $\gf(q)$.  Schmidt and White have given a characterization of all two-weight
irreducible cyclic codes over $\gf(q)$ when $q$ is prime  \cite{SW}. However, the
conditions for the characterization given in \cite{SW} cannot be easily used for finding
out all all two-weight irreducible cyclic codes over $\gf(p)$. It follows from (\ref{eqn-forall1})
that the code $\C(r,N)$ in (\ref{eqn-irrecode}) has at most two nonzero weights if and
only if the Gaussian periods $\eta_i^{(\gcd((r-1)/(q-1),N),r)}$  take on at most two distinct
values. A special case of this is the case of uniform cyclotomy \cite{BMW}. It might be
possible to give another chacaterization in this direction.

\section{The weight distribution in a few other cases and other results}

Gaussian periods of order 5, 6, 8 and 12 are computed in \cite{Hosh} and \cite{Gura} respectively. So
the weight distribution of the code $\C(r,N)$ in (\ref{eqn-irrecode}) can be computed by these Gaussian
periods and  (\ref{eqn-wtmain}). However, the weight formulas will be complicated due to the messy
expression of these Gaussian periods.  Two-weight projective irreducible cyclic codes are characterized
by Wolfmann \cite{Wolf}.

Two recursive algorithms were developed for computing the weight distribution of
certain irreducible cyclic codes \cite{Moisio}.  The weight enumerators of all nondegenerate
irreducible cyclic binary [$n$ , $m$]-codes have been computed for which $k > 27$ and
$N = (2^m - 1)/n < 500$ by Ward \cite{Ward}. The weights of irreducible cyclic
codes are discussed by Aubry and Langevin \cite{AL}, Moisio \cite{Mois} and by Segal and Ward \cite{SWa}.
The relations between the weight
distributions of irreducible cyclic codes and the Hasse-Davenport curves are dealt with
by van der Vlugt \cite{Vlug}.  Chains of irreducible cyclic codes and relations among their
weight distributions are presented in \cite{Klov,HKM}.

\section{Bounds on weights in irreducible cyclic codes}

Since it is notoriously hard to determine the weight distribtions of the irrreducible cyclic codes,
it would be interesting to develop tight bounds on the weights in   irrreducible cyclic codes. Such
tight bounds can give information on the error-correcting capability of this class of cyclic codes.
The objective of this section is to develop such tight bounds.

\begin{theorem}\label{thm-bounds}
Let $N$ be a positive divisor of $r-1$ and define  $N_1=\gcd((r-1)/(q-1), N)$.  Let $m_0$ be
the nultiplicative order of $q$ modulo $n$. Then the set
$\C(r,N)$ in (\ref{eqn-irrecode}) is a $[(q^m-1)/N, m_0]$ cyclic code over $\gf(q)$ in which the weight
$w$ of every nonzero codeword satisfies that
\begin{eqnarray*}
w_H(c(\beta)) &\ge& (q-1) \left\lceil \frac{r - \lfloor(N_1-1)\sqrt{r}\rfloor}{qN} \right\rceil, \\
w_H(c(\beta)) &\le& (q-1)\left\lfloor \frac{r + \lfloor(N_1-1)\sqrt{r}\rfloor}{qN}    \right\rfloor.
\end{eqnarray*}
In particular, if $N_1(N_1-1)<r$, then $m_0=m$.
\end{theorem}

\begin{proof}
The results of this theorem follow from Theorem \ref{thm-wdivisibility} and (\ref{eqn-wtmain}).
\end{proof}

The lower bound of Theorem \ref{thm-bounds} is tight when $\gcd((r-1)/(q-1),N)$ is small,
and may not be tight in some other cases. When $\gcd((r-1)/(q-1),N)=1$, the lower and
upper bounds of Theorem \ref{thm-bounds} are the same, and they are indeed achieved
as the code in this case is a constant-weight code. Table \ref{tab-bounds} lists some
experimental data, where $n$, $k$, $d$ are the length, dimension and minimum nonzero
weight of the code.

\vspace{0.1cm}
\begin{table}[ht]
\caption{The lower bound of Theorem \ref{thm-bounds}}\label{tab-bounds}
\begin{center}
{\begin{tabular}{|c|c|c|c|c|c|} \hline
$n$ & $k$ & $d$ & $q$ & lower bound of Thm \ref{thm-bounds}  & $\frac{r-1}{q-1} \bmod N$\\
\hline \hline
5   & 4    &  2  & $2$   & 2  &     0 \\ \hline
21   & 6    & 8   & $2$   & 8   &   0 \\ \hline
21   & 3    & 12   & $2^2$   & 12   &   0 \\ \hline
85   & 4    & 64   & $2^2$   & 64  &   1\\ \hline
13   & 3    & 9   & $3$   & 9  &   1\\ \hline
40   &   4  &  24  & $3$   &  24  &   0 \\ \hline
121   &  5   &  81  & $3$   &  81 &   1 \\  \hline
312   & 4    & 240   & $5$   & 236  &  0 \\ \hline
\end{tabular}
}
\end{center}
\end{table}

\section{Summery and open problems}

The contributions of this paper include the following:
\begin{itemize}
\item A survey of earlier results on the weight distributions of irreducible cyclic codes.
\item Extensions and generalizations of earlier results on the weight distributions of irreducible cyclic codes
(Theorems \ref{thm-semipri}, \ref{thm-quadra}, \ref{thm-2wt0}, \ref{thm-2wt1}, \ref{thm-2wt1111},
\ref{thm-e137}, and \ref{thm-ccc4}).
\item A complete characterization of one-weight irreducible cyclic codes (Theorem \ref{thm-1wt2}), which is an extension of the result in \cite{Vega}.
\item The weight divisibility of irreducible cyclic codes (Theorem \ref{thm-gaussianp}).
\item A lower and upper bound on the weights in irreducible cyclic codes (Theorem \ref{thm-bounds}).
\item A property on Gaussian periods (Theorem \ref{thm-wdivisibility})
\end{itemize}

While it is hard to determine the weight distributions of the irreducible cyclic codes in general,
it is possible to solve this problem for other special cases. One open problem would be a simpler
characterization of two-weight irreducible cyclic codes than the one presented in \cite{SW} by
Schmidt and White.

\end{document}